\documentclass[aps,pra,reprint]{revtex4-2}

\usepackage[utf8]{inputenc}

\usepackage{amsmath}
\usepackage{amssymb}
\usepackage{amsfonts}
\usepackage{graphicx} 
\usepackage{hyperref} 
\usepackage{xcolor}
\usepackage[normalem]{ulem}

\begin{document}

\title{Phase-Space Engineering and Collective Dynamics in Memcomputing}

\author{Chesson Sipling}
\email{email: csipling@ucsd.edu}
\affiliation{Department of Physics, University of California San Diego, La Jolla, CA 92093}

\author{Yuan-Hang Zhang}
\email{email: yuz092@ucsd.edu}
\affiliation{Department of Physics, University of California San Diego, La Jolla, CA 92093}

\author{Massimiliano Di Ventra}
\email{email: diventra@physics.ucsd.edu}
\affiliation{Department of Physics, University of California San Diego, La Jolla, CA 92093}

\begin{abstract}
Digital Memcomputing machines (DMMs) are dynamical systems with memory (time non-locality) that have been designed to solve combinatorial optimization problems. Their corresponding ordinary differential equations depend on a few hyper-parameters that define both the system's relevant time scales and its phase-space geometry. Using numerical simulations on a prototypical DMM, we analyze the role of these physical parameters in engineering the phase space to either help or hinder the solution search by DMMs. We find that the DMM explores its phase space efficiently for a wide range of parameters, aided by the system-wide correlations in their fast degrees of freedom that emerge {\it dynamically} due to coupling with the (slow) memory degrees of freedom. In this regime, the time it takes for the system to find a solution scales well as the number of variables increases. When these hyper-parameters are chosen poorly, the system navigates its phase space far less efficiently. However, we find that, in many cases, collective behavior persists even when the phase-space exploration process is inefficient. This behavior only disappears if the memories are made to evolve as quickly as the fast degrees of freedom. This study points to the important role of memory and hyper-parameters in engineering the DMMs' phase space for optimal computational efficiency.
\end{abstract}

\maketitle

\section{Introduction}
Memcomputing represents a new type of computing paradigm where {\it memory} is employed as a computing tool~\cite{diventra13a,traversa2015universal,di2022memcomputing}. Here, ``memory'' refers to the general time non-local nature of the corresponding dynamical system (rather than the more restrictive meaning of ``data storage''). This memory plays a crucial role in enabling memcomputing machines to solve challenging computational problems efficiently. These machines can be fabricated using a variety of materials and devices, including electrical devices, optical elements, etc.~\cite{di2022memcomputing}, so long as a time non-local response is present in the dynamical system~\cite{Membook}. Starting from these experimental realizations one can then write down the corresponding nonlinear ordinary differential equations (ODEs) describing their dynamics. In turn, these ODEs can be simulated on traditional computers even without a hardware realization.

Within the general class of memcomputing machines, {\it digital} memcomputing machines (DMMs), map a finite string of symbols into a finite string of symbols, enabling them to solve combinatorial optimization problems~\cite{traversa2017polynomial,di2018perspective}. Their internal ODEs couple fast degrees of freedom (DOFs) --representing the variables of the problem to solve-- to slow, memory DOFs. The memory variables enable the fast DOFs to become {\it strongly coupled} during the solution search; the resulting collective behavior leverages the global structure of the problem to find a logical solution efficiently~\cite{sipling2024memory}. In addition, the DMMs' phase space can be engineered so that the only attractors of the dynamics are the solutions of the problem to solve, and all other critical points are saddle points~\cite{di2022memcomputing}. Altogether, this makes a DMM a powerful tool when tackling difficult combinatorial optimization problems.


In this work, we study numerically and provide some analytical understanding of the dynamics of prototypical DMMs, concentrating on the way changing hyper-parameters influences their ability to solve a given problem. The main findings of our work are as follows:


1. There exists a {\it wide} range of viable DMM hyper-parameters where the solution search is efficient. This {\it phase} of solvability is related to the existence of a region in phase space in which the fast DOFs evolve {\it collectively}. This behavior is induced by memory. The fact that a {\it phase} exists where solutions can be found efficiently indicates that DMMs are robust against small perturbations of the hyper-parameters within this range.


2. Despite there being a wide range of hyper-parameters that yield collective dynamics, only a subset of these parameters is accompanied by scale-invariance. While collective behavior is certainly required to navigate highly non-convex landscapes efficiently, scale-invariance does not seem to be always necessary.

3. When the memory DOFs vary over time scales comparable to those of the fast DOFs, the DMM's ability to solve the problem decreases dramatically. This loss of efficiency originates from the breakdown of collective behavior in the fast DOFs, as timescale separation is required to generate system-wide correlations dynamically.

4. On the opposite end, when the memory dynamics are extremely slow, the DMM's performance worsens as well. We attribute this to the ``flattening'' of the unstable manifolds of the saddle points, which, in the presence of numerical noise, favors the existence of time-reversed trajectories (i.e., anti-instantons, in a field theory language). These time-reversed trajectories are instead exponentially suppressed in the efficient-solution phase.


5. In agreement with previous studies~\cite{PhysRevE.108.034306}, the distribution of time-to-solution (TTS) is fitted well by an inverse Gaussian distribution. Interestingly, we detect such distributions even when the parameters are chosen ``poorly''. This suggests that the overall mechanism for DMMs to traverse their phase space --consisting of quick ``jumps'' between saddle points of increasing stability~\cite{di2022memcomputing}-- remains structurally unchanged even when the hyper-parameters are altered significantly. Instead, we expect that a dramatic change in the {\it topology} of the phase space is required to alter this characteristic behavior~\footnote{Note that such a topological change may also occur due to the introduction of ``ghost critical points'' by higher-order numerical integration schemes~\cite{stuart1994numerical}.}.

The paper is organized as follows. In Sec.~\ref{memory_role}, we analyze the role that memory plays in DMMs, both establishing collective dynamics in the fast DOFs and shaping/navigating the DMM phase space. In Sec.~\ref{DMM3SAT}, we introduce the 3-SAT problem and the full DMM equations which we simulate to solve hard 3-SAT instances. Finally, in Sec.~\ref{Numres}, we perform some numerical analyses to substantiate the claims made above.

\section{Memory's role in memcomputing} \label{memory_role}

\subsection{Collective dynamics via memory}

We begin with a general analysis of some generic DMMs' equations of motion that does not depend on their precise functional form. In fact, apart from the necessity of having two types of DOFs, slow (memory) and fast (variables of the problem to solve), the ODEs of DMMs may be different even for a specified computational problem~\cite{di2022memcomputing}. That said, we will now provide analytical arguments on how, within a certain region in hyper-parameter space, the fast DOFs behave collectively due to the presence of memory, which aids the system in the solution search process.

Let's then consider the following generic DMM dynamics, which feature a set of variables $v_n \in [-1, 1]$ (representing the logical variables of the problem the DMM is designed to solve) coupled to memory variables $x_m \in [0, 1]$:

\begin{align}\label{eqn:simple_v_dot}
    \dot{v}_n &= \sum_m \Big( x_m A_{n,m}(\vec{v}) + B_{n,m}(\vec{v}) \Big), \\
    \dot{x}_m &= \beta (C_m(\vec{v}) - \gamma), \label{eqn:simple_x_dot}
\end{align}

\noindent where $n = 1, 2, ..., N$ and $m = 1, 2, ..., M$, with $N$ the number of logical variables and $M$ the number of memory variables. Both $C_m(\vec{v}) \in [0, 1]$ and $\gamma \in [0, 1]$.

Note that all variables have been {\it continuously relaxed}; e.g., $v_i \in \{-1, 1\} \rightarrow [-1, 1]$ (where we take $-1$ and $+1$ to be the booleans FALSE and TRUE, respectively, in the original, logical problem). For the relaxed variables, it is enough to assign $v_i>0$ to the logical TRUE, and $v_i<0$ to the logical FALSE. This relaxation is crucial in order to transform the discrete configuration space of the original, logical problem into a continuous phase space on which our DMM can evolve. In Sec.~\ref{DMM3SAT} we will provide explicit functional forms for the functions $A_{n,m}(\vec{v})$, $B_{n,m}(\vec{v})$, and $C_m(\vec{v})$. Here, we only assume them to be continuous real-valued functions on the space of $v_i$'s.

We will now show how the $v_i$'s can evolve collectively in such a system due to the memory variables $x_m$'s. Similar calculations have been performed recently in the context of neuromorphic devices~\cite{zhang2024collective}, spin systems~\cite{sipling2024memory}, and the biological brain~\cite{sun2024memory}. This process requires that the memory variables evolve more slowly than the logical variables so that the $x_m$'s preserve some ``memory'' of the system's recent history. In fact, the presence of such memory is necessary in order for DMMs to function properly.

To infer the existence of collective behavior, we iteratively integrate our dynamics, similar to how a computer would simulate such dynamics numerically. A similar technique has been used in previous works in other contexts, including transport theory~\cite{10.1063/1.1724117} and neural dynamics~\cite{wilson_cowan}. We find this approach to be much more effective (and feasible) than, e.g., multiple-scale analysis~\cite{17039} or any other existing semi-analytical techniques. During the iterative integration, we fix the integration timestep to be $\tau_x$, the characteristic timescale of the memory variable dynamics, which satisfies $\tau_x > \tau_v$, with $\tau_v$ as the characteristic timescale of the logical variable dynamics.

While we cannot identify $\tau_v$ so easily from Eq.~(\ref{eqn:simple_v_dot}) without knowing the actual form of $A_{n,m}(\vec{v})$ and $B_{n,m}(\vec{v})$, we readily see from Eq.~(\ref{eqn:simple_x_dot}) that $\tau_x \approx 1/\beta$ (since $\gamma$ and $C_m(\vec{v})$ are generally $O(1)$~\cite{di2022memcomputing}). Therefore, after one timestep $\tau_x$,

\begin{align} \label{eqn:v_1tau}
    v_n(\tau_x) &= v_n(0) + \int_0^{\tau_x} dt \, \dot{v}_n(t) \\
    &\approx v_n(0) + \frac{1}{\beta} \sum_m \Bigg(x_m(0) \overline{A_{n,m}^{(0)}(\vec{v})} + \overline{B_{n,m}^{(0)}(\vec{v})} \Bigg),
\end{align}

\noindent where we have defined

\begin{equation} \label{eqn:avg}
    \overline{F^{(l)}(\vec{v})} \equiv \frac{1}{r} \sum_{a=0}^{r-1}F\big(\vec{v}(l \tau_x + a \tau_v)\big),
\end{equation}

\noindent with $F^{(l)}(\vec{v})$ either $A_{n,m}(\vec{v})$ or $B_{n,m}(\vec{v})$, $l=0, 1, \dots$, and $r \equiv \lfloor \tau_x / \tau_v \rfloor$, with $\lfloor \cdot \rfloor$ the floor function of the argument. Essentially, Eq.~(\ref{eqn:avg}) gives the average of some function $F$ in the time interval $[l\tau_x, (l+1)\tau_x)$ of width $\tau_x$. Notice that $x_m$ does not change much during time intervals shorter than $\tau_x$, so we can treat it as approximately constant during each integration step. Also, we must have $r \gg 1$ for the system to have memory.

Similarly, for the memory variables,

\begin{align} \label{eqn:x_1tau}
    x_m(\tau_x) &= x_m(0) + \int_0^{\tau_x} dt \, \dot{x}_m(t) \\
    &\approx x_m(0) + \frac{1}{\beta} \Bigg( \beta \overline{C_m^{(0)}(\vec{v})} - \gamma \Bigg).
\end{align}

Now, we integrate Eq.~(\ref{eqn:simple_v_dot}) from $\tau_x$ to $2 \tau_x$:

\begin{align} \label{eqn:x_2tau}
    v_n(2 \tau_x) &= v_n(\tau_x) + \int_{\tau_x}^{2 \tau_x} dt \, \dot{v}_n(t) \\
    &\approx v_n(\tau_x) + \frac{1}{\beta} \sum_m \Bigg( x_m(\tau_x) \overline{A_{n,m}^{(1)}(\vec{v})} + \overline{B_{n,m}^{(1)}(\vec{v})} \Bigg) \\
    &\approx \frac{1}{\beta} \sum_m \overline{C_m^{(0)} (\vec{v})} \, \overline{A_{n,m}^{(1)}(\vec{v})} + \dots. \label{eqn:last_line}
\end{align}

In Eq.~(\ref{eqn:last_line}), we have written only the most relevant term for our analysis out of a series of terms. Assuming that $C_m$ and $A_{n,m}$ depend non-trivially on some $v_i \in \vec{v}$ (with $i \neq n$), this term couples $v_n$ to other logical variables in the system non-linearly. As this iterative approach is continued, additional nonlinear couplings between logical variables like this will continue to appear, and the couplings will become less and less local in the sense that more and more $v_i$'s will influence the evolution of $v_n$.

Taken alone, this is not so surprising. One would expect high-order, non-local couplings to emerge as an iterative integration approach like this is performed by hand. One might also 
expect these terms to act only as small perturbations to the state of a logical variable $v_n$ at a given time, so that the evolution of $v_n$ depends primarily upon only a few $v_i$'s. However, crucially, if $\beta \lesssim |\sum_m \overline{C_m^{(0)} (\vec{v})} \, \overline{A_{n,m}^{(1)} (\vec{v})}|$, this effect is {\it non-perturbative}. So, the logical variable $v_n$ will instead become {\it strongly} influenced by many $v_i$'s, up to as many as there are in the entire system. As this occurs for each variable simultaneously, the entire system becomes strongly coupled. Thus, the presence of memory can induce collective behavior among the relatively fast DOFs.

Note that this analysis requires that $\beta$ not be too large. If instead $\beta \gtrsim |\sum_m \overline{C_m^{(0)} (\vec{v})} \, \overline{A_{n,m}^{(1)} (\vec{v})}|$, the effect becomes {\it perturbative} (so that many $v_i$'s do not correlate strongly nor evolve collectively) precisely when $x_m$ evolves too quickly (so it is no longer an effective source of memory on the system's recent history). Of course, the exact value of $\beta$ which qualifies as ``too large'' will depend on the functional form of $A_{n,m}(\vec{v})$ and any physical parameters therein (cf. Sec.~\ref{DMM3SAT}). In the Supplemental Material (SM), we elaborate upon this analysis further with a simple example.

\subsection{Effective phase space navigation via memory}

The dynamical, collective effect we have just described is precisely what enables DMMs to solve combinatorial optimization problems efficiently. These problems are typically plagued by highly non-convex landscapes in their state spaces, and traditional algorithms (which rely only on ``local'' information in the solution search process) frequently get stuck in local minima~\cite{complexity_book}. For example, a traditional algorithm that quickly finds a reasonably good configuration (i.e., a low-energy configuration that is not the ground state) may still need to change the value of many (a macroscopic number of) variables to reach the true solution (the ground state). It can then be quite challenging to find a better configuration by updating variables one or a few at a time. Instead, by introducing memory into the system, the (continuously relaxed) logical variables in DMMs become strongly coupled during the solution search so that they can evolve {\it collectively} (leveraging the global structure of the problem) towards a logical solution.


These memory variables also help by “opening up” directions in the phase space, allowing the system to escape from regions that would otherwise be local minima in the subspace of logical variables $v_i$'s alone. In this expanded phase space, composed of both logical and memory DOFs, points that were local minima in the original problem can now exhibit non-zero dynamics along the ``memory directions''. This means that, although these points may remain stationary in the logical subspace, the additional memory dynamics can guide the system away from them, enabling continued progress toward a solution.

Additional insight into the collective dynamics of DMMs has been obtained by means of supersymmetric topological field theory~\cite{di2017topological, di2019digital}, showing that the only low-energy/long-wavelength trajectories in the DMMs' phase space are a collection of successive {\it instantons}. Practically speaking, these instantons correspond to sudden ``jumps'' in phase space in which many logical variables (even as many as the size of the problem) change suddenly. Instantons are then the underlying expression of collective behavior in DMMs. Importantly, these jumps are between saddle points of decreasing index (with the index being the number of unstable directions of the critical point), until the system eventually converges to a solution (with index $0$). During each jump, the system's state is constrained to lie on an instanton manifold. The union of these manifolds and the critical points between them (collectively called the ``composite instanton'') is a proper subset of the entire phase space (in fact it is a CW complex~\cite{di2022memcomputing}). Therefore, during this characteristic DMM search process, the machine breaks ergodicity. Finally, the dynamics will only stop once a solution (if it exists) is reached (which is a fixed point of the flow field). This behavior has been explicitly shown in several previous studies~\cite{di2017topological, PhysRevE.100.053311, di2022memcomputing}.

However, as we have just discussed, this characteristic DMM search process relies upon the proper choice of hyper-parameters. As we will demonstrate shortly, poorly chosen parameters can dramatically change the effectiveness of a DMM at finding a solution to a given problem. This can occur due to a breakdown of collective behavior in the fast DOFs and/or the influence of strong numerical noise~\cite{zhang2021directed}. In fact, a similar limitation in the presence of strong physical noise has been recently shown in Ref.~\cite{Nguyen2025}.


\section{DMM dynamics for 3-SAT} \label{DMM3SAT}

\subsection{The 3-SAT problem}

Since our goal is to understand the role of hyper-parameters in the solution search of DMMs, we focus on 3-SAT instances with planted solutions, so there is no ambiguity in whether a solution exists or not. A single instance of the 3-SAT problem is characterized by a set of $N$ logical variables and $M$ clauses, where each clause contains 3 literals (a literal being a logical variable or its negation). Each clause is TRUE if at least one of its literals is TRUE—that is, the literals are connected by a logical OR. The full 3-SAT problem is a logical AND of all these clauses, so a solution must satisfy every clause at the same time.

The instances with planted solutions have been generated according to the procedure outlined in Ref.~\cite{barthel2002hiding}. This approach was used in previous SAT-solving competitions~\cite{balyo2016using, heule2018generating} and is known to be challenging for traditional algorithms (i.e., it requires an exponentially increasing number of steps to solve as $N$ increases~\cite{barthel2002hiding}). We fix the clause-to-variable ratio, $\alpha_r \equiv M/N$, to $4.3$, which lies close to the satisfiable-to-unsatisfiable phase transition and is known to yield some of the hardest instances~\cite{CRAWFORD199631}.

\subsection{DMM's 3-SAT equations}

We now introduce one possible set of DMM equations that we will simulate numerically, which have proven effective (in the correct parameter regime) in solving the above 3-SAT instances~\cite{bearden2020efficient}~\footnote{Note that the equations used in the present work are slightly different from those used in Ref.~\cite{bearden2020efficient}, in the representation of the short- and long-term memory variables (see the SM for more details). These changes do not affect the solution search, further reinforcing that DMMs are quite robust even with respect to structural changes of their flow vector field~\cite{di2022memcomputing}.}:

\begin{widetext}
\begin{align}
    \dot{v}_n &= \sum_m x_m^l x_m^sG_{n,m}(v_n, v_j, v_k) + \sum_m(1 + \zeta x_m^l)(1 - x_m^s)R_{n,m}(v_n, v_j, v_k), \label{eqn:DMM1} \\
    \dot{x}_m^s &= \beta(C_m(v_i,v_j,v_k) - \gamma), \label{eqn:DMM2} \\
    \dot{x}_m^l &= \begin{cases} \alpha(C_m(v_i,v_j,v_k)-\delta), &C_m(v_i,v_j,v_k)\geq\delta, \\ \alpha(C_m(v_i,v_j,v_k)-\delta)(x_{l,m}-1) , &C_m(v_i,v_j,v_k)<\delta, \label{eqn:DMM3} \end{cases} \\
    G_{n,m}(v_n, v_j, v_k) &= \frac{1}{2} q_{n,m} \min[(1-q_{j,m}v_j),(1-q_{k,m}v_k)], \label{eqn:DMM4} \\
    R_{n,m}(v_n,v_j,v_k) &= \begin{cases} \frac{1}{2}q_{n,m}(1-q_{n,m}v_n), &C_m(v_n,v_j,v_k) = \frac{1}{2}(1-q_{n,m}v_n), \\ 0, &\text{else}, \end{cases} \label{eqn:DMM5} \\
    C_m(v_i, v_j, v_k) &= \frac{1}{2}\min[(1-q_{i,m}v_i),(1-q_{j,m}v_j),(1-q_{k,m}v_k)]. \label{eqn:DMM6}
\end{align}
\end{widetext}

$v_i, v_j,$ and $v_k$ are the 
three variables of a single 3-SAT OR gate, each of which evolves according to Eq.~(\ref{eqn:DMM1}). $x^s_m$ and $x^l_m$ are the short- and long-term memory DOFs, which increase or decrease depending on the state of the corresponding clause function (only one of each exists for each clause). While not necessary to the memcomputing paradigm itself, previous studies~\cite{bearden2020efficient} have found that having two types of memory DOFs with the above functional forms aids in the solution search. Finally, $\alpha$, $\beta$, $\gamma$, $\delta$, and $\zeta$ are the physical hyper-parameters. $\beta$ and $\alpha$ influence the rate of change of the short- and long-term memories, respectively (with $\alpha < \beta$). $\gamma$ and $\delta$ are thresholding parameters that must both be less than $1/2$. $\zeta$ influences the degree of ``rigidity'' in the system, or more precisely, the long-term memories' relative influence on the system's rigidity (see Ref.~\cite{bearden2020efficient} for more discussion on the role of these quantities).

Note that the 3-SAT instance is solved if $C_m<1/2$ for all clauses~\cite{bearden2020efficient}. Once this occurs, the simulation is stopped, and the sign of the logical variables can be read as a boolean string that corresponds to the problem's logical solution. When simulating these equations, each instance is initialized with random $v_i = \pm 1$ and fixed $x_m^s = 1/2$ and $x_m^l = 1$. These equations correspond to Eq.~(\ref{eqn:simple_v_dot}) and Eq.~(\ref{eqn:simple_x_dot}), with $A_{n,m}$, $B_{n,m}$, and $C_m$ defined as follows:

\begin{align}
    A_{n,m}(\vec{v}) &= x_m^lG_{n,m}(v_n, v_j, v_k) - B_{n,m}(\vec{v}), \\
    B_{n,m}(\vec{v}) &= (1 + \zeta x_m^l) R_{n,m}(v_n, v_j, v_k), \\
    C_m(\vec{v}) &= C_m(v_i, v_j, v_k).
\end{align}

Notice that in Eqns.~(\ref{eqn:DMM4}),~(\ref{eqn:DMM5}), and~(\ref{eqn:DMM6}), $q_{i, m}$ indicates whether or not a logical variable is negated in the literal $l_{i, m}$ appearing in the relevant 3-SAT formula: $q_{i, m} = 1$ if $l_{i, m} = v_i$ (in clause $m$), and  $q_{i ,m} = -1$ if $l_{i, m} = \lnot v_i$. Since our solution search process is {\it dynamical} rather than algorithmic, to make a particular literal $l_{i, m}$ TRUE in the original discrete problem, we must ``push'' $v_i$ towards $q_{i, m}$. Thus, we can also reformulate any literal $l_{i, m}$ as a continuous dynamical variable $q_{i, m} v_{i, m}$. See the SM for a more detailed description of these dynamics.


We emphasize that this is just one possible form of DMM's dynamics—the specific functional form is not unique, as long as there is timescale separation between the logical DOFs and at least one type of memory DOF. In fact, different ODE formulations may be more effective for different types of problems~\footnote{Note also that in order to accelerate numerical simulations, the parameters in Eqns.~(\ref{eqn:DMM1}),~(\ref{eqn:DMM2}), and~(\ref{eqn:DMM3}) have been chosen so that the dynamics approach critical points in the phase space but never ``fall into'' one exactly (except for the equilibrium points). This speeds up the approach to equilibrium without qualitatively changing the phase space exploration.}.

\begin{figure*}
    \centering
    \includegraphics[width=\textwidth]{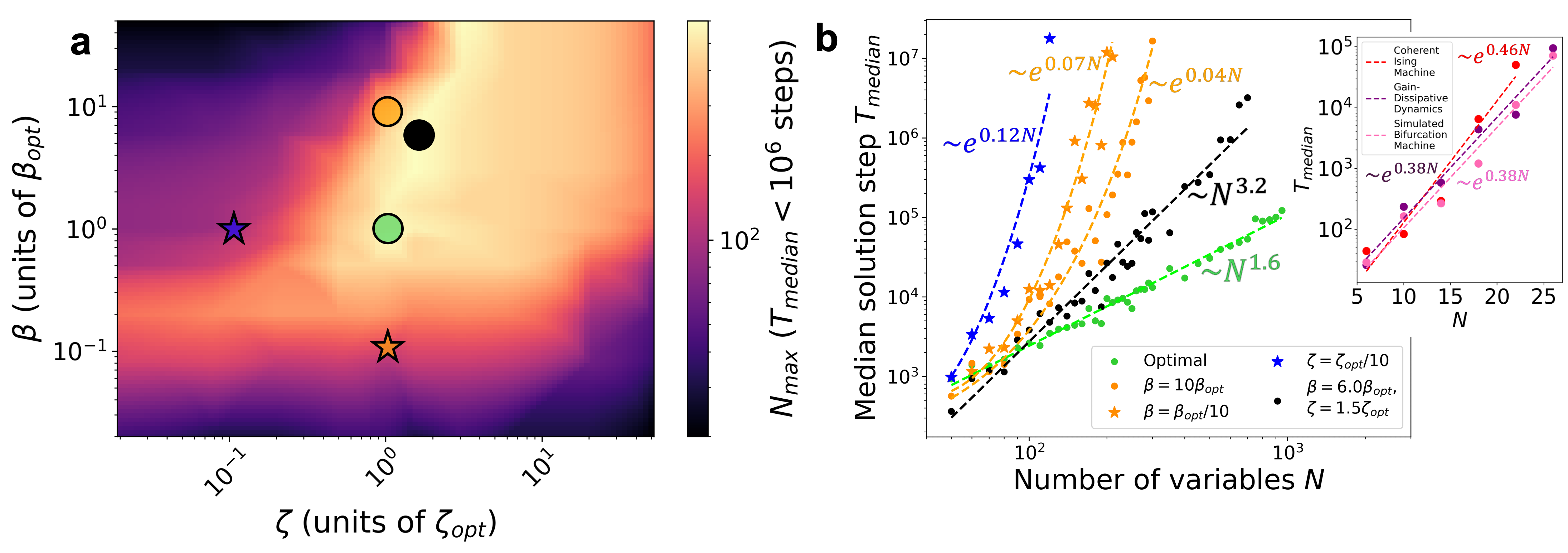}
    \caption{(a) State space diagram in hyper-parameter space $\{\beta, \zeta\}$, showing the maximum size 3-SAT ($\alpha_r = 4.3$) instance $N_{max}$ that a DMM can solve within a fixed number of integration steps $T_{median}$ with fixed $dt_0 = 1.0$. The number of logical variables, $N$, is increased by $10$ at a time until $T_{median}$ exceeds $10^6$ steps. We find a wide region in phase space in which moderately large $N$ instances ($\sim400-600$) can be solved quickly without tuning $dt_0$ (which scales the adaptive time step). Colored dots and stars correspond to parameter choices depicted in (b). (b) Scalability plots of the median number of steps to find a solution $T_{median}$ as a function of number of logical variables $N$. For two sets of parameters ($\{ \beta, \zeta \} =  \{ \beta_{opt}, \zeta_{opt} \}= \{ 79, 0.063\} $ and $\{ 6.0 \beta_{opt}, 1.5 \zeta_{opt} \}$), the $T_{median}(N)$'s are  fitted well by polynomials ($\sim N^{1.65}$ and $\sim N^{3.18}$, respectively). Outside the phase of ``good'' parameters, $T_{median}(N)$'s are fitted well by exponentials ($\sim e^{0.12N}$, $\sim e^{0.07N}$, and $\sim e^{0.04N}$ for $\{ \beta, \zeta \} =  \{ \beta_{opt}, \zeta_{opt}/10 \}$, $\{ \beta_{opt}/10, \zeta_{opt} \}$, and $\{ 10\beta_{opt}, \zeta_{opt} \}$, respectively). Here, $dt_0$ is tuned at each data point to control numerical errors; everywhere, it takes a value between $0.01$ and $1.0$. In comparison with the memcomputing approach, the inset shows that $T_{median}$ scales exponentially for three different physics-inspired solvers, including coherent Ising machines~\cite{tiunov2019annealing}, gain-dissipative dynamics~\cite{kalinin2018global}, and simulated bifurcation machines~\cite{goto2016bifurcation}. In both figures, batches of $100$ instances are run in parallel, and the batch terminates when at least half of all instances are solved. $\{\alpha/\beta, \beta, \gamma, \delta, \zeta \}_{opt} = \{0.45, 79, 0.36, 0.080, 0.063 \}$ in both (a) and (b).}
    \label{fig:Fig1}
\end{figure*}

Although Eqns.~(\ref{eqn:DMM1}),~(\ref{eqn:DMM2}), and~(\ref{eqn:DMM3}) are ``stiff'' (they involve two distinct time scales), due to the topological nature of the solution search by DMMs, explicit integration methods have been found to suffice~\cite{di2022memcomputing, zhang2021directed}. Although surprising at first glance, this is because the exact trajectory the DMM takes in phase space is irrelevant, so long as the numerical noise is not so strong as to push the trajectory off its composite instanton manifold~\cite{di2017topological, DIVENTRA2019167935, bearden2020efficient}. We will then simulate these ODEs using the simplest method: a forward Euler integration scheme with an adaptive time step $dt = {dt_0}/({\max{\{|\dot{v_i}|\}}} + \epsilon)$ (clamped after evaluation so that $dt \leq 10^{-1}$). $\epsilon$ (set to $10^{-6}$) prevents $dt$ from diverging when $\max{\{|\dot{v_i}|\}} = 0$.

It must be emphasized that there is a crucial difference between the {\it physical} parameters in Eqns.~(\ref{eqn:DMM1}),~(\ref{eqn:DMM2}), and~(\ref{eqn:DMM3}) and the {\it numerical} parameter $dt_0$. Although some $dt_0$ tuning is required to control numerical errors, only $\alpha$, $\beta$, $\gamma$, $\delta$, and $\zeta$ influence physical DMM properties such as the geometry of the phase space and the correlations between logical variables. For example, if a DMM was constructed in {\it hardware}~\cite{zhang2025implementation, nguyen2023hardware, nguyen2024fully} that evolved according to Eqns.~(\ref{eqn:DMM1}) through~(\ref{eqn:DMM6}), $dt_0$ would not play any role at all, as there is no longer any notion of an ``integrated trajectory'' which approximates the true one. In all other ways, physical noise is expected to influence the dynamical behavior of a memcomputing machine built in hardware in the same ways as numerical noise~\cite{Nguyen2025}. A key example in which either numerical or physical noise could play the same role will be discussed in Sec.~\ref{anti-inst}.

Since we are most interested in the {\it physical} properties of DMMs in this work, we will concentrate only on the {\it physical} DMM parameters for the remainder of this manuscript. 



\section{Numerical results} \label{Numres}

We are now ready to study in detail the relevance of the different physical hyper-parameters on the DMM's ability to solve instances. In this study, we fix $\gamma$, $\delta$, and the relative value of $\alpha/\beta$. We then exclusively concentrate on the 2D parameter space $\{\beta, \zeta\}$, as these two parameters appear to influence the phase space dynamics most strongly.

As a starting point, we have used a parallel tempering (PT) approach~\cite{earl2005parallel} to initialize the parameters, focusing on 3-SAT instances up to $N = 10^2$. We have found the following optimal parameters (in terms of time to solution) at that size: $\{\alpha/\beta, \beta, \gamma, \delta, \zeta\}_{opt} = \{0.45, 79, 0.36, 0.080, 0.063\}$. See the SM for more details on our initial parameter-tuning procedure.

Although we will refer to these parameters as ``optimal'' for the remainder of this work, a more advanced tuning procedure, or a more sophisticated metric, could potentially yield an even better set of parameters. This is especially true when considering other problem sizes much smaller or larger than $N = 10^2$. The numerator in our adaptive time step, $dt_0$, will act as a 6th numerical parameter which will be used exclusively to control numerical errors.

\subsection{Range of viable hyper-parameters}

Now that we have tuned and thereby fixed the initial values of $\alpha/\beta$, $\gamma$, and $\delta$, we can explore the 2D parameter space populated by the key parameters $\beta$ and $\zeta$.

First, we test the quality of a wide range of pairs $\{ \beta, \zeta \}$ in enabling DMMs to solve large 3-SAT instances (see Fig.~\ref{fig:Fig1}). Batches of $100$ instances are run in parallel until either a median number of instances are solved, or $10^6$ simulation timesteps are exceeded. Fig.~\ref{fig:Fig1}(a) reveals a wide region in parameter space in which moderately large instances could be solved in a relatively short amount of time. In fact, $\beta$ and $\zeta$ can each be varied by a factor of at least 2-3 in either direction without changing the DMM's efficiency in solving the problem too significantly.

In Fig.~\ref{fig:Fig1}(b), a few of these parameters are studied in more detail to extract their scalabilities (in median number of steps to solution, $T_{median}$, as a function of $N$). Since these simulations are pushed to a larger number of steps ($\sim 10^7$), $dt_0$ is tuned at each system size $N$ via PT to control numerical errors. Interestingly, for the highest quality parameters, $T_{median}(N)$ can be fitted well by polynomials, allowing us to solve many of the hardest ($\alpha_r = 4.3$) 3-SAT instances within $10^5$ time steps. Note that the choice of $\{ \beta, \zeta \} =  \{ 6.0 \beta_{opt}, 1.5 \zeta_{opt} \}$ yields a $T_{median}$ which is also fitted well by a polynomial. On the other hand, further from optimality (lying in darker regions in Fig.~\ref{fig:Fig1}(a)), $T_{median}$'s are fitted well by exponentials, making solving large-$N$ instances untenable with such parameters. Due to our limited computational resources, we were not able to perform simulations with tuned $dt_0$ at all tested points in parameter space.

To contextualize these results, we compare against three modern physics-inspired Ising solvers: coherent Ising machines~\cite{tiunov2019annealing}, gain-dissipative dynamics~\cite{kalinin2018global}, and simulated bifurcation machines~\cite{goto2016bifurcation}. These methods optimize an Ising (quadratic unconstrained binary optimization) objective and therefore require a SAT-to-Ising mapping~\cite{lucas2014ising, anthony2017quadratic}. On the planted 3-SAT benchmarks considered here, they exhibit exponential time-to-solution scaling, and none were able to solve instances with $N>30$ within $10^5$ steps. Comparison with other SAT solvers, such as WalkSAT and Survey Inspired Decimation were carried out previously in Ref.~\cite{bearden2020efficient}, showing again DMMs' superior success probabilities and scaling behavior across system sizes.

\begin{figure*}[t]
    \centering
    \includegraphics[width=\textwidth]{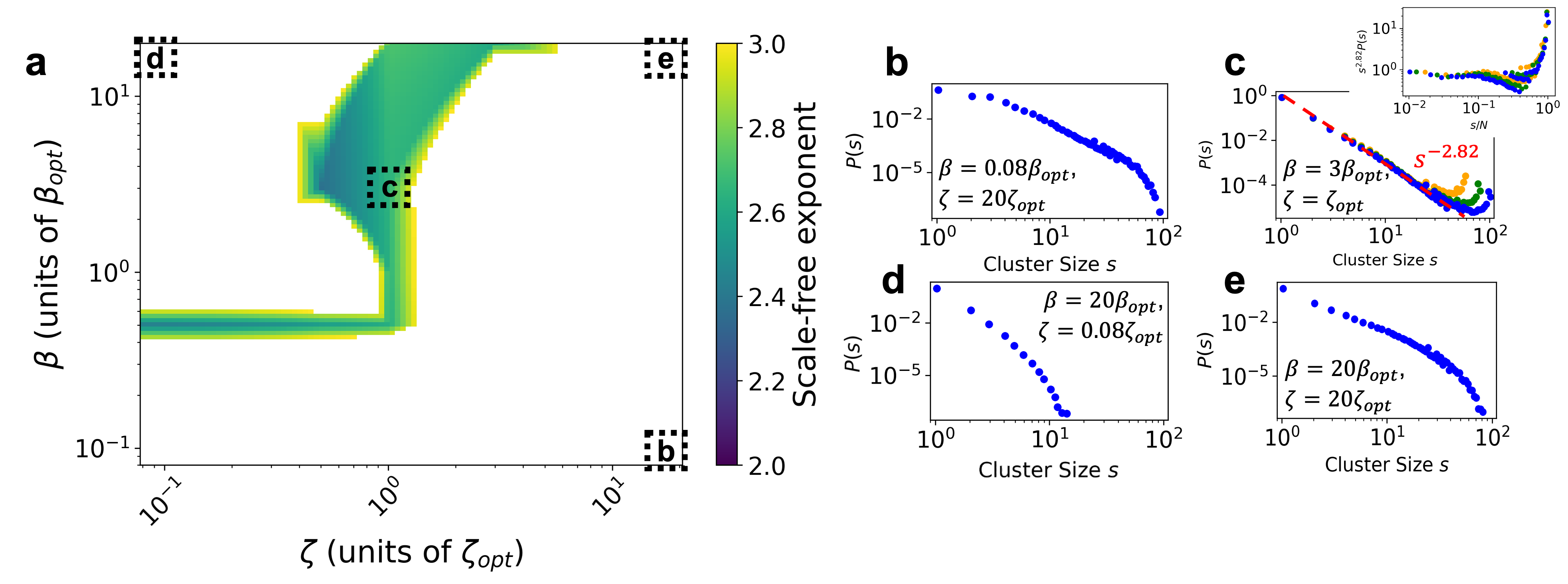}
    \caption{(a) Diagram in $\{ \beta, \zeta \}$ (in units of $\{ \beta_{opt}, \zeta_{opt} \}= \{ 79, 0.063\} $) characterizing the degree of scale-freeness of avalanche distributions, alongside (b)-(e) four of these avalanche distributions at four points in parameter space, labeled in panel (a). 
    In diagram (a), a parameter range exists where scale-free (SF) avalanche distributions are detected with an exponent between $2$ and $3$. Of the four points in the diagram (a) whose distributions are shown, three are not SF (points {\bf b}, {\bf d}, and {\bf e}). However, some of these distributions (at points {\bf b} and {\bf e}) still feature large avalanches that approach the system size. At point {\bf c}, the distribution is fitted by $s^{-2.82}$ and includes a characteristic bump due to finite size effects, with finite size scaling plotted in the inset. At point {\bf d}, only small avalanches are detected, as the lack of memory in the system destroys the collective behavior. 100 batches are simulated, and each batch includes 100 instances run for $T = 1000$ steps with $N=100$ (except for (c), where $N=60$ is plotted in orange, and $N=80$ is plotted in green alongside $N=100$ in blue). The time window to detect the avalanches is $\Delta_{tw} = 1.0$, $0.006$, $0.0006$, and $0.035$ for points {\bf b}, {\bf c}, {\bf d}, and {\bf e}, respectively (see the SM for more details).} 
    \label{fig:Fig2}
\end{figure*}

\subsection{Collective dynamics, but not necessarily scale-invariance}


As anticipated in Sec.~\ref{memory_role}, collective behavior is the key physical property that enables DMMs to efficiently find a solution to a given problem. To corroborate this, we study the emergence of correlations in DMMs by extracting avalanche distributions of the fast DOFs. We define an avalanche as follows. When the boolean value of any $v_i$ changes, we say an event has occurred at time $t$ and location $i$. We then search for similar flipping events at nearest-neighbors of $v_i$, defined as the other logical variables that appear in the same clauses as $v_i$, within a window of time. If one of the nearest-neighbors of $v_i$ flips before this window of time has passed, the size of the avalanche is increased by 1, and further events are then searched for adjacent to the newest event. See the SM for more details regarding the generation of these avalanches. These can be visualized as ``waves'' that propagate through the system comprised of many $v_i$'s whose corresponding boolean values flip rapidly. In our phase space dynamics, avalanches correspond to jumps along instantons.

In Fig.~\ref{fig:Fig2}, we plot the associated scale-free exponents, if they exist, of these avalanche distributions at many points in our $\{ \beta, \zeta \}$ parameter space. In a wide region roughly centered at $\{ \beta_{opt}, \zeta_{opt} \} = \{79, 0.063\}$, we detect scale-free (SF) distributions with exponents between $2$ and $3$. One such distribution, at $\{ \beta, \zeta \} = \{ 3 \beta_{opt}, \zeta_{opt} \}$ (point {\bf c} in Fig.~\ref{fig:Fig2}(a)), is plotted in panel Fig.~\ref{fig:Fig2}(c). In its inset, we perform a finite size scaling analysis according to the finite size scaling ansatz $P(s) \sim s^{-\alpha^\prime}\exp(-s/N)$~\cite{finite_size_scaling_Fisher}, where $\alpha^\prime$ is the fitted power law exponent. The fact that the distributions at sizes $N=60$, $80$, and $100$ collapse after imposing this ansatz suggests these avalanche distributions' deviations from power laws (in particular, the finite size ``bumps'') can be attributed to finite size effects, and that these distributions do approach power laws in the thermodynamic limit.

Interestingly, at most other points in parameter space, avalanche distributions still feature large (up to the system size $N$) avalanches. This is not too surprising since many parameters were already shown in Fig.~\ref{fig:Fig1}(a) to be viable, and when $\beta$ is low, we expect parameters to be poor due to the prevalence of time-reversed trajectories (anti-instantons), not a lack of collective behavior (cf. Sec.~\ref{anti-inst}). Examples include $\{ \beta, \zeta \} = \{ 0.08 \beta_{opt}, 20 \zeta_{opt} \}$ and $\{ 20 \beta_{opt}, 20 \zeta_{opt} \}$ (points {\bf b} and {\bf e} in Fig.~\ref{fig:Fig2}(a), respectively), both of which are plotted Fig.~\ref{fig:Fig2}(b) and (e). This suggests that, even if DMMs do not possess perfect scale-invariance for some parameter choices, their memories are nonetheless quite effective at inducing system-wide correlations in the logical variables. It is only due to the strong coupling between $v_i$'s (i.e., their {\it collective} behavior) due to memory, that large clusters of $v_i$'s can all ``flip'' their logical values in short succession, corresponding to such large avalanches.

We expect that, in the thermodynamic limit, the subset of viable parameters (the bright, central region in Fig.~\ref{fig:Fig1}(a)) should converge to the parameter regime that yields SF avalanche distributions. As evidence, we note the direct overlap between the black and green dots in Fig.~\ref{fig:Fig1}(a) (whose $T_{median}(N)$ distributions are fitted well by polynomials) and the region of scale-freeness in Fig.~\ref{fig:Fig2}(a). Furthermore, consider the distributions in  Fig.~\ref{fig:Fig2}(b) and (e). While they do feature large avalanches approaching size $N = 10^2$, their correlations decay faster than the SF case at point {\bf c} (Fig.~\ref{fig:Fig2}(c)), and they showcase no finite size ``bump''. Therefore, if we were to perform simulations with larger $N$ at points {\bf b} and {\bf e}, we would not expect the size of the largest avalanches to increase much beyond $10^2$. As $N$ is increased further, arbitrarily large correlated clusters are required to solve instances efficiently, meaning SF-ness would become necessary in the large-$N$ limit. Unfortunately, our computational resources prevent us from extracting avalanche distributions for larger instances with reasonable statistics.

\subsection{Loss of collective behavior when memories are too fast}

Crucially, point {\bf d} ($\{ \beta, \zeta \} = \{ 20 \beta_{opt}, 0.08 \zeta_{opt} \}$) in Fig.~\ref{fig:Fig2}(d)) {\it does not} feature any large-scale avalanches. The reason is quite easy to understand. From our general analysis in Sec.~\ref{memory_role}, we found that $\beta$ cannot be too large if collective dynamics are to be established, which is necessary for the DMM to capitalize on the global structure of the problem. Therefore, we anticipated that once $\beta$ exceeded a particular value, memory-induced collective behavior should break down and the DMM would have to rely on purely ``local'' information.

However, recall from that analysis that we needed $\beta \lesssim |\sum_m \overline{C_m^{(0)}(\vec{v})} \, \overline{A_{n,m}^{(1)}(\vec{v})}|$ in order for $v_i$ to evolve collectively, which implicitly depends on $A_{n,m}(\vec{v}) = x_m^lG_{n,m}(v_n, v_j, v_k) - (1 + \zeta x_m^l) R_{n,m}(v_n, v_j, v_k)$. Therefore, by varying $\zeta$, we also vary the timescale associated with the {\it logical} DOFs. This leads to the ``stabilizing'' effect present in Fig.~\ref{fig:Fig1}(a); when $\beta$ and $\zeta$ are large {\it simultaneously}, timescale separation between the logical and memory DOFs, and thus collective behavior, persists. This effect is less pronounced when $\beta$ and $\zeta$ are instead both small, as the term $(1 + \zeta x_m^l) R_{n,m}$ helps to bound the logical DOF timescale from above. This is precisely what we observe numerically; when $\beta$ is too large (and $\zeta$ does not accordingly increase to offset it), the timescale separation between the memory and logical DOFs is destroyed, eliminating the collective dynamics in the fast DOFs. Here, it corresponds to a lack of large-scale avalanches.


\subsection{Anti-instantons when memories are too slow}\label{anti-inst}

\begin{figure}
    \centering
    \includegraphics[width=\columnwidth]{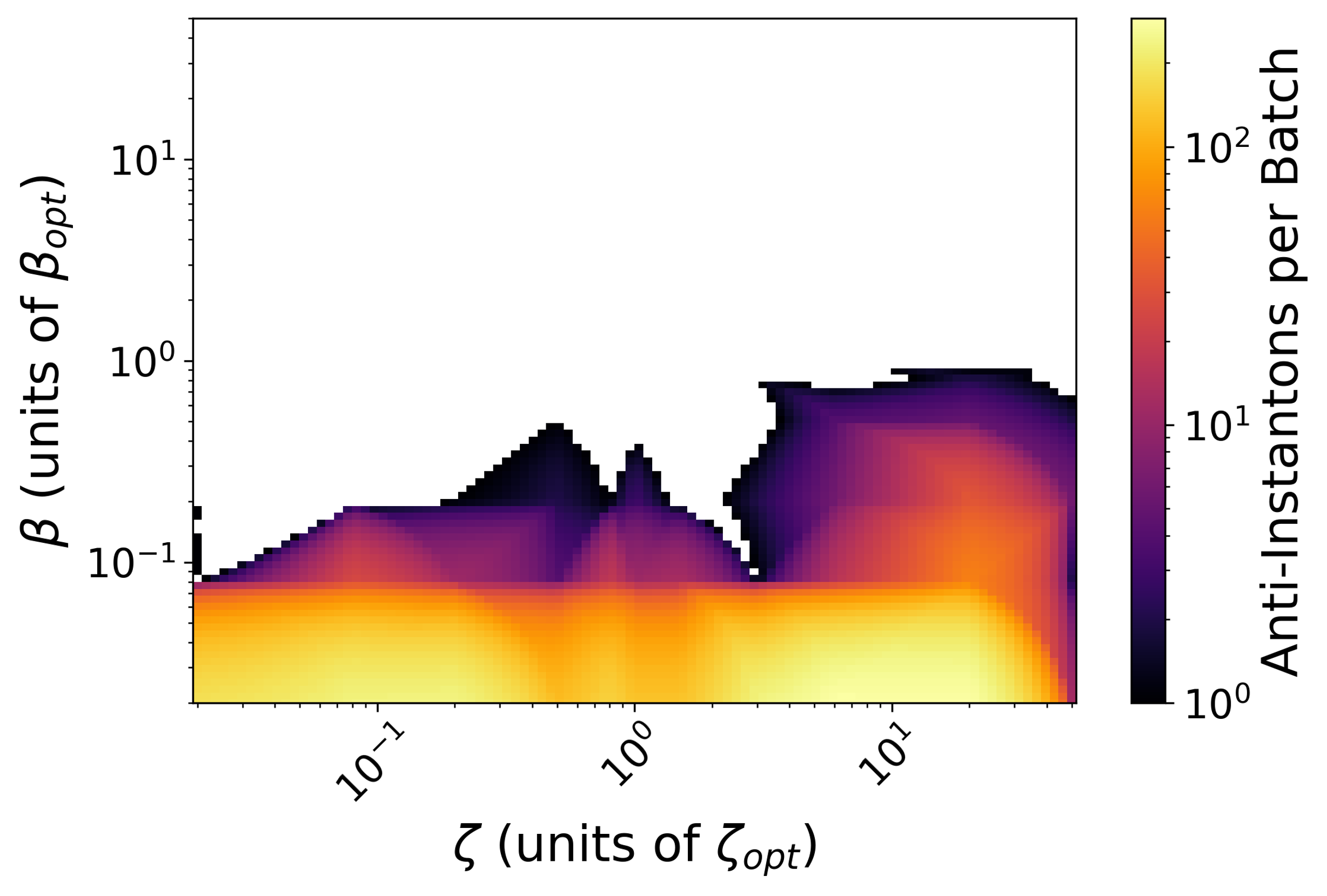}
    \caption{Diagram in $\{\beta, \zeta\}$ (in units of $\{ \beta_{opt}, \zeta_{opt} \}= \{ 79, 0.063\} $) featuring the number of time-reverse trajectories (anti-instantons) present during the DMM solution search (per batch). Anti-instantons are detected by extracting a time-ordered list of avalanches in $v_i$'s (collections of flips within a short time interval) and keeping track of when avalanches consisting of identical sets of $v_i$'s occur in immediate succession (see the SM for details). We only detect anti-instantons when $\beta$ is low. This region overlaps significantly with the dark region at the top of Fig.~\ref{fig:Fig1}(a). 15 batches are simulated, each batch includes 100 instances each with $N = 100$, and simulations are stopped after 6000 time steps.}
    \label{fig:Fig3}
\end{figure}

Another region of parameter space where the DMM does not solve the instances efficiently is evident in Fig.~\ref{fig:Fig1}(a) and corresponds to small $\beta$. Here, an entirely different phenomenon occurs: time-reverse trajectories ({\it anti-instantons}) emerge due to the presence of numerical noise~\cite{di2022memcomputing}. We can understand this phenomenon as follows.

During the DMM's solution search, instantons (or avalanches) are characterized by a large collection of logical variables $v_i$'s flipping sign (i.e., changing their corresponding boolean value) in quick succession. From the perspective of phase space navigation, these instantonic ``jumps'' drive the trajectory closer and closer to the global fixed point, which is also the logical solution.

The corresponding time-reversed processes (also known as anti-instantons) correspond to a collection of flipping $v_i$'s which instead drive the system {\it away} from the solution (i.e., towards a saddle with {\it larger} index). Typically, this process is exponentially suppressed~\cite{di2022memcomputing}, but it can occur more often when the noise (whether physical or numerical) is large enough to overcome the ``energetic'' barrier for the system to transition from a critical point with given stability to a {\it less} stable one. This is similar to water flowing downhill: the water cannot flow uphill unless strong noise is introduced into the system.

Numerically, we can track all the $v_i$'s that flip in each instantonic jump. If we ever detect the {\it same} collection of $v_i$'s flipping in back-to-back instantons, an anti-instanton must have occurred. For example, if we detect instantons where $\{v_1, v_4, v_5\}$ flip, then $\{v_2, v_4, v_7\}$, and then $\{v_2, v_4, v_7\}$ again, we can be sure that at least one anti-instanton occurred. There may be more anti-instantons that cannot be detected via this approach. However, at the very least, it provides a lower bound on the number of anti-instantons generated during dynamics.

We plot the number of anti-instantons detected as a function of $\beta$ and $\zeta$ in Fig.~\ref{fig:Fig3}. They only occur when $\beta$ is small. Small $\beta$ corresponds to very slowly varying memory DOFs, meaning there is a much lower degree of curvature in the ``memory directions'' in phase space. So, the numerical errors that accumulate in the solution search will be more problematic in this regime, since the separation of ``energy levels'' between saddles on either side of an instanton can become quite small. Therefore, only a small amount of numerical noise is required to push the system against the flow that leads to the solution. In the analogy of water flowing downhill, it takes considerably less noise for the water to occasionally flow uphill, the smaller the slope of the hill. In the limit of zero slope, the water has the same probability of flowing ``downhill'' as ``uphill'' in the presence of noise.

Note that an explicit calculation of the curvature of the stable and unstable manifolds near saddles points is computationally intractable for multi-dimensional phase spaces. Therefore, we find this method of extracting anti-instantons to be more efficient and sufficient evidence to describe the poor performance of DMMs in the small-$\beta$ regime.

Also, note the strong overlap in phase space between the region with anti-instantons in Fig.~\ref{fig:Fig3} and the dark region of poor parameters in Fig.~\ref{fig:Fig1}(a). This suggests that these anti-instantons, arising due to the increased influence of numerical noise accompanying small $\beta$, are likely responsible for the DMM's exceptionally poor performance in that region of parameter space.

\subsection{Time-to-solution distributions}

\begin{figure}
    \centering
    \includegraphics[width=\columnwidth]{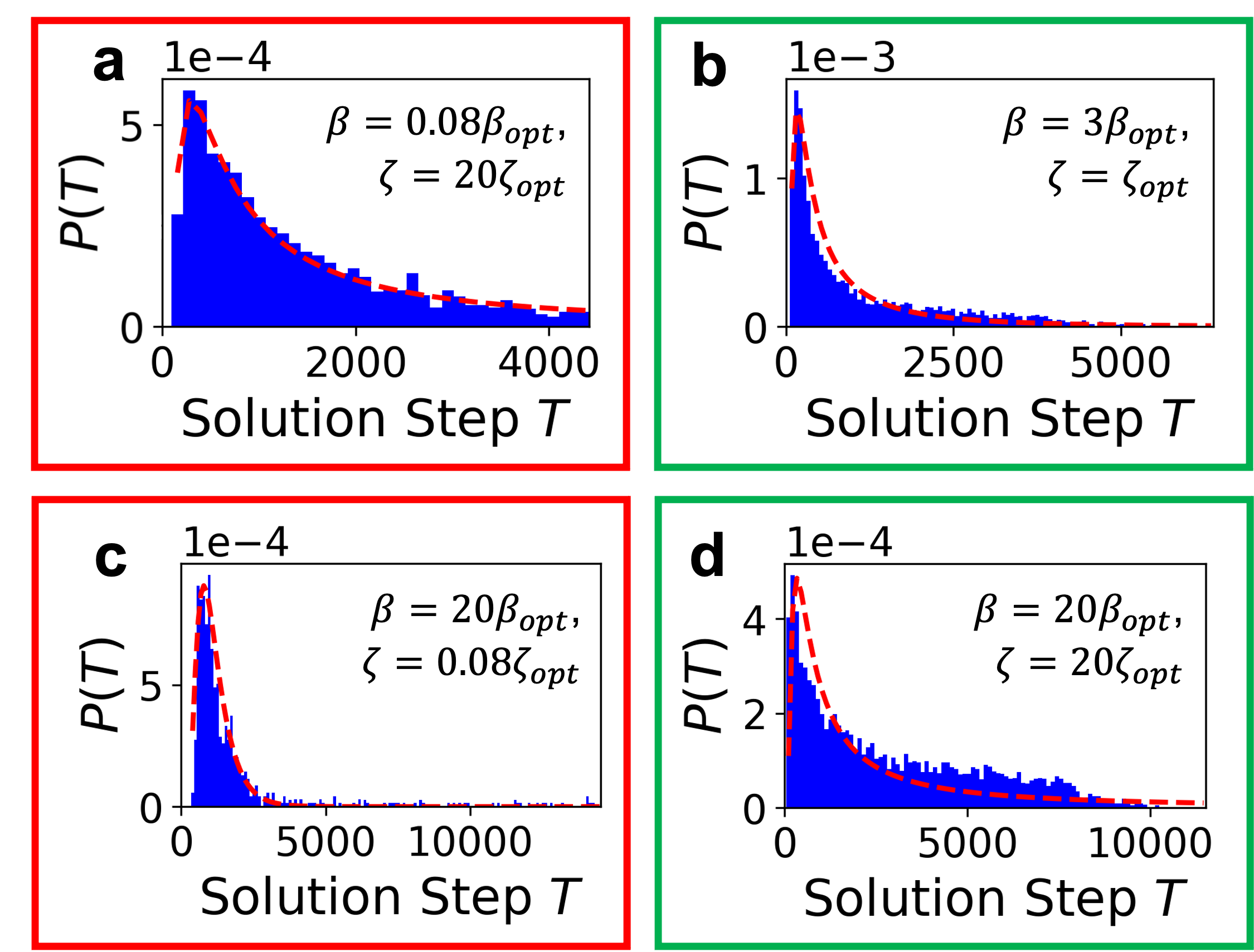}
    \caption{A comparison of the time-to-solution (TTS) distributions $P(T)$ as a function of number of solution steps $T$ for different $\{ \beta, \zeta \}$. Plots (a), (b), (c), and (d) correspond to plots (b), (c), (d), and (e) in Fig.~\ref{fig:Fig2}. The distributions in all plots are fitted to inverse Gaussians. 100 batches are simulated, and each batch includes 100 instances each with $N = 100$. According to the results of Fig.~\ref{fig:Fig1}(a), plots (a) and (c) correspond to parameters that lead to inefficient solution search, while the parameters in (b) and (d) lead to an efficient search. For (a), the median solution step is $T_{median} \approx 10^7$. For (c), a median of instances could not be solved within $T = 10^8$ steps.}
    \label{fig:Fig4}
\end{figure}


Finally, we investigate the distributions of time-to-solution (TTS) of DMMs for the same $\{ \beta, \zeta \}$ pairs as those reported in Fig.~\ref{fig:Fig2}. Previous work~\cite{PhysRevE.108.034306} has already shown that for parameter values that lead to an efficient solution search, DMMs' TTS distributions are {\it inverse Gaussian} (IG). The reason for this type of distribution is because the DMM's phase space navigation in the presence of (perturbative) noise (whether physical or numerical) is analogous to that of a driven Brownian particle, where the ``particle position'' in the DMM case corresponds to the vector of $v_i$'s.


In Fig.~\ref{fig:Fig4}, TTS distributions are plotted at the four points in $\{ \beta, \zeta \}$ parameter space featured in Fig.~\ref{fig:Fig2}(a). In Fig.~\ref{fig:Fig4}(a) and (c) ($\{ \beta, \zeta \} = \{ 0.08\beta_{opt}, 20\zeta_{opt} \}$ and $\{ 20\beta_{opt}, 0.08\zeta_{opt} \}$, respectively), the parameters correspond to ``poor'' performance of the DMM (see Fig.~\ref{fig:Fig1}(a)). In Fig.~\ref{fig:Fig4}(b) and (d) ($\{ \beta, \zeta \} = \{ 3\beta_{opt}, \zeta_{opt} \}$ and $\{ 20\beta_{opt}, 20\zeta_{opt} \}$), they correspond to the efficient search by the DMM.

Interestingly, at all points in parameter space, even when parameters lead to poor performance, the TTS distributions are fitted reasonably well by IGs. The fact that distributions are still IGs even in this case, indicates that the DMMs' characteristic phase space dynamics are largely unchanged. Both in the case of $\{ \beta, \zeta \} = \{ 0.08\beta_{opt}, 20\zeta_{opt} \}$ (Fig.~\ref{fig:Fig4}(a)), where we detect the most anti-instantons due to numerical noise, and $\{ \beta, \zeta \} = \{ 20\beta_{opt}, 0.08\zeta_{opt} \}$ (Fig.~\ref{fig:Fig4}(c)), where the lack of timescale separation prevents the logical variables from evolving collectively, the distributions have this same characteristic shape. The distribution in Fig.~\ref{fig:Fig4}(d) does deviate slightly from a pure inverse Gaussian, but we attribute this to the significant increase in phase space curvature at that point in parameter space, which is a purely {\it geometric} change to the phase space. This suggests that the DMM's phase space topology remains unaffected even after varying parameters away from optimality by more than an order of magnitude.


In other words, although the lack of collective behavior and the existence of anti-instantons certainly hinder a DMM's solution search, neither seem to significantly push the system off the composite instanton manifold. When there is no longer collective behavior in the fast DOFs, some instantonic trajectories (those corresponding to many logical DOFs flipping) become inaccessible, so a longer path in phase space must be taken to reach the equilibrium. When there are anti-instantons, the system may reverse its path along an instanton, nullifying some previous progress towards the equilibrium. Importantly, in neither case does the system need to navigate its entire $N$-dimensional phase space (ergodicity is still broken). This clarifies why these difficult instances can be solved at all even with such a poor choice of parameters (albeit with correspondingly poor scalability as a function of problem size).

To observe something other than an IG TTS distribution, one would have to break the phase space topology entirely by, {\it e.g.}, introducing additional critical points (in particular, additional local minima which {\it do not} correspond to logical solutions). It is then not surprising that, for many initial conditions, a DMM with such a phase space would become unable to ever reach a logical solution.

\section{Conclusions}

Collective dynamics in fast DOFs induced by memory (slow DOFs) has already been shown to exist in many other physical contexts, from spin systems~\cite{sipling2024memory} to neuromorphic devices~\cite{zhang2024collective} to models of human cortical dynamics~\cite{sun2024memory}. DMMs are yet another example of dynamical systems with two distinct DOFs in which such a phenomenon emerges and is further leveraged to solve combinatorial optimization problems.

In this work, we have performed an extensive study of the role of the hyper-parameters entering the ODEs of DMMs to engineer the DMMs' phase space and affect the collective evolution of the logical variables. We find a wide range of parameters to be ``viable'', meaning that the DMMs with those parameters are able 
to reach equilibrium (solution of the problem to solve) quite efficiently.


When there is no longer sufficient timescale separation between the two DOFs, the system no longer evolves collectively and the solution search by DMMs becomes inefficient. Practically, this corresponds to the hyper-parameter $\beta$ being too large (without $\zeta$ being increased accordingly, which would reintroduce timescale separation) so that the memory variables evolve too rapidly. A similar inefficiency occurs when the memory DOFs are extremely slow. However, in this case, the origin of such inefficiency is related to the emergence of (numerical) noise-induced time-reversed trajectories (anti-instantons), which occasionally reverse the flow toward the solution, hindering the search process. In our simulations, this occurs when $\beta$ is too small. We expect a similar phenomenon to emerge for the hardware version of these machines when they are subject to physical noise.

Although only a subset of viable parameters provide scale invariance (as suggested by the avalanche distributions), the latter does not seem to be always necessary to solve problem instances up to the sizes we could reach with our computational resources. Many other regions in parameter space feature large avalanches up to the largest problem size, suggesting that strong coupling among the logical DOFs (and thus, collective dynamics) still exists even without such scale invariance. However, further (analytical) work is needed to understand if the phase of collective behavior will eventually ``merge'' with the phase of scale invariance in the thermodynamic limit, and what the consequences are for the efficiency of DMMs to solve problem instances in such a limit.


In addition to shedding further light on the physics behind these machines, this work paves the way for more efficient methods to tune the relevant hyper-parameters of the ODEs of DMMs. In fact, we anticipate that such tuning procedures would have to rely on maximizing a metric which assesses the collective behavior in the fast DOFs, which is the most important physical advantage of these machines. We leave this for future studies.\\

\section*{Acknowledgments}
We thank Val\'erian Demaurex and Fabio L. Traversa for useful discussions. C.S., Y.-H. Z., and M. D. are supported by the NSF grant No. ECCS-2229880.

\section*{Author contributions}
M.D. suggested and supervised the work. C.S. performed the theoretical analysis and numerical simulations. Y.-H.Z. developed the software code used to execute the numerical simulations. All authors have read and contributed to the writing of the paper.


\section*{Data, materials, and software availability}

The Python simulation code for data generation is available at~\cite{github_link}.

\bibliography{apssamp}

\end{document}


\title{Supplemental Material for ``Phase-Space Engineering and Collective Dynamics in Memcomputing''}

\author{Chesson Sipling}
\email{email: csipling@ucsd.edu}
\affiliation{Department of Physics, University of California San Diego, La Jolla, CA 92093}

\author{Yuan-Hang Zhang}
\email{email: yuz092@ucsd.edu}
\affiliation{Department of Physics, University of California San Diego, La Jolla, CA 92093}

\author{Massimiliano Di Ventra}
\email{email: diventra@physics.ucsd.edu}
\affiliation{Department of Physics, University of California San Diego, La Jolla, CA 92093}

\maketitle

\section{Supplementary Analyses}

\subsection{A concrete example of the emergence of collective behavior via memory}

Recall that in the main text, we defined some generic DMM dynamics with the following equations:

\begin{align}\label{eqn:simple_v_dot}
    \dot{v}_n &= \sum_m \Big( x_m A_{n,m}(\vec{v}) + B_{n,m}(\vec{v}) \Big), \\
    \dot{x}_m &= \beta (C_m(\vec{v}) - \gamma). \label{eqn:simple_x_dot}
\end{align}

We then proceeded to perform an iterative integration approach to explain how the logical variable $v_n$ could become strongly coupled to other $v_i$'s in the system, enabling the entire system to evolve {\it collectively} after some time had passed. Here, we would like to show the emergence of such strong coupling more explicitly with a concrete example.

For simplicity, let's consider a 1D lattice of $N$ evenly spaced, continuously relaxed logical variables $v_i \in [-1, 1]$, with $i=1,2,... N$ and periodic boundary conditions. Each variable only interacts explicitly with its neighbor to the left, $v_{i-1}$. Additionally, there is a single memory variable $x_i$, defined on the bonds between $v_i$ and $v_{i-1}$, which grows when $v_i$ and $v_{i-1}$ are not aligned and shrinks when they are aligned. The interactions encourage each $v_i$ to align with its left-neighbor, so long as $x_i$ is sufficiently large (they have been anti-aligned for a sufficiently long amount of time). Otherwise, $v_i$ is held at its current value. We can then define $A_n = v_{n-1} - v_n$, $B_n = v_n$, and $C_m = -v_{n-1}v_n + \gamma$. Upon substituting these expressions into Eq.~(\ref{eqn:simple_v_dot}) and~(\ref{eqn:simple_x_dot}), we obtain:


\begin{align}
    \dot{v}_n &= x_n (v_{n-1} - v_n) + v_n, \label{eqn:v_dot_example} \\
    \dot{x}_n &= -\beta v_{n-1}v_n . \label{eqn:x_dot_example}
\end{align}

We can again confirm that the memory timescale $\tau_x$ satisfies $\tau_x \approx 1/\beta$, since $v_{n-1}$ and $v_n$ are both $O(1)$. In this specific example, since $x_n$ must also be $O(1)$, we can further see that $\tau_v$ must satisfy $\tau_v \approx 1$.

Let's next perform the first few integration steps of width $\tau_x$ explicitly, as in the main text. Once again, since $x_n$ does not change much over time intervals shorter than $\tau_x$, we can treat it as approximately constant over each integration step.

\begin{widetext}
\begin{align}
    v_n(\tau_x) &= v_n(0) + \int_0^{\tau_x} dt \, \dot{v}_n(t) \\
    &\approx v_n(0) + \frac{1}{\beta} \Bigg( x_n(0) \bigg( \overline{v^{(0)}_{n-1}} - \overline{v^{(0)}_n} \bigg) + \overline{v^{(0)}_n} \Bigg).
\end{align}

\begin{align}
    x_n(\tau_x) &= x_n(0) + \int_0^{\tau_x} dt \, \dot{x}_n(t) \\
    &\approx x_n(0) - \overline{v_{n-1}v_n^{(0)}}.
\end{align}

\begin{align}
    v_n(2\tau_x) &= v_n(\tau_x) + \int_{\tau_x}^{2\tau_x} dt \, \dot{v}_n(t) \\
    &\approx v_n(\tau_x) + \frac{1}{\beta} \Bigg( x_n(\tau_x) \bigg( \overline{v^{(1)}_{n-1}} - \overline{v^{(1)}_n} \bigg) + \overline{v^{(1)}_n} \Bigg) \\
    &\approx v_n(0) + \frac{1}{\beta} \Bigg( x_n(0) \bigg( \overline{v^{(0)}_{n-1}} - \overline{v^{(0)}_n} \bigg) + \Big( x_n(0) - \overline{v_{n-1}v_n^{(0)}} \Big) \bigg( \overline{v^{(1)}_{n-1}} - \overline{v^{(1)}_n} \bigg) + \overline{v^{(0)}_n} + \overline{v^{(1)}_n} \Bigg).
\end{align}

\begin{align}
    x_n(2 \tau_x) &= x_n(\tau_x) + \int_{\tau_x}^{2\tau_x} dt \, \dot{x}_n(t) \\
    &\approx x_n(\tau_x) - \overline{v_{n-1}v_n^{(1)}} \\
    &\approx x_n(0) - \overline{v_{n-1}v_n^{(1)}} - \overline{v_{n-1}v_n^{(0)}} .
\end{align}
\begin{align}
    v_n(3 \tau_x) &= v_n(2\tau_x) + \int_{2\tau_x}^{3\tau_x} dt \, \dot{v}_n(t) \\
    &\approx v_n(2\tau_x) + \frac{1}{\beta} \Bigg( x_n(2\tau_x) \bigg( \overline{v^{(2)}_{n-1}} - \overline{v^{(2)}_n} \bigg) + \overline{v^{(2)}_n} \Bigg) \\
    &\approx v_n(0) + \frac{1}{\beta} \Bigg( x_n(0) \bigg( \overline{v^{(0)}_{n-1}} - \overline{v^{(0)}_n} \bigg) + \Big( x_n(0) - \overline{v_{n-1}v_n^{(0)}} \Big) \bigg( \overline{v^{(1)}_{n-1}} - \overline{v^{(1)}_n} \bigg) \\
    &\qquad\qquad + \Big( x_n(0) - \overline{v_{n-1}v_n^{(1)}} - \overline{v_{n-1}v_n^{(0)}} \Big) \bigg( \overline{v^{(2)}_{n-1}} - \overline{v^{(2)}_n} \bigg) + \overline{v^{(0)}_n} + \overline{v^{(1)}_n} + \overline{v^{(2)}_n} \Bigg) .
\end{align}
\end{widetext}

Above, we note a pattern beginning to emerge. At later integration steps, high-order nonlinear couplings begin to emerge which implicitly couple $v_n$ to other variables further in the lattice. The most relevant terms are those of the form $-\frac{1}{\beta} \overline{v_{n-1}v_n^{(p)}}\bigg( \overline{v_{n-1}^{(p+1)}} - \overline{v_n^{(p+1)}} \bigg)$, which is precisely the $\frac{1}{\beta} \overline{C_m^{(0)}} \, \overline{A_{n,m}^{(1)}}$ term referenced in the main text with $p=0$. After a much larger number of integration steps, since these functions of $v_{n-1}$ and $v_n$ are being averaged up to later times, they will implicitly depend on other $v_{n-k}$ further away in the lattice. When $\beta$ is sufficiently small to act as a source of memory ($\beta \ll 1$ in this case, since $\tau_v \approx 1$), this implicit dependence is significant compared to other variables which are closer in the lattice. This strongly couples the $v_n$ in the system, enabling them to evolve {\it collectively}.

In this example, we can imagine how the system might evolve dynamically. Let's initialize the system so $x_i(0) = 0$ and $v_i(0) = \pm 1$, assigned randomly from a uniform probability distribution, for all $i$. Initially, no variables will flip, as Eq.~(\ref{eqn:v_dot_example}) holds $v_i$ at its existing value when $x_i$ is small (recall that $v_i$ is bounded between $-1$ and $1$). However, since $v_i$ are assigned randomly, a variety of $(v_{i-1}, v_i)$ pairs will initially be anti-aligned. At those sites, memory will slowly accumulate until it is sufficient to induce a flip in $v_i$. Many such $v_i$'s will flip in quick succession, constituting a global event where a macroscopic number of variables change rapidly. These events will continue to occur as time passes and subsequent ``avalanches'' alter the distribution of variables $v_i$'s until a ground state is found (with either all $v_i = -1$ or all $v_i = 1$).

These macroscopic readjustments where {\it many} logical variables flip at nearly the same time would not be possible without a source of memory. If we had $\beta \gtrsim 1$, individual $v_i$'s which are anti-aligned with their left neighbor would flip almost immediately, which would not give each $v_i$ enough time to correlate strongly with other variables far away in the lattice. This is exactly the same phenomenon we observe in the full DMM equations in the main text.

\section{DMM equations}

\subsection{Details of DMM dynamics}

For convenience, we will repeat here the full DMM equations reported in the main text:\\

\begin{widetext}
\begin{align}
    \dot{v}_n &= \sum_m x_m^l x_m^sG_{n,m}(v_n, v_j, v_k) + \sum_m(1 + \zeta x_m^l)(1 - x_m^s)R_{n,m}(v_n, v_j, v_k), \label{eqn:DMM1} \\
    \dot{x}_m^s &= \beta(C_m(v_i,v_j,v_k) - \gamma), \label{eqn:DMM2} \\
    \dot{x}_m^l &= \begin{cases} \alpha(C_m(v_i,v_j,v_k)-\delta), &C_m(v_i,v_j,v_k)\geq\delta, \\ \alpha(C_m(v_i,v_j,v_k)-\delta)(x_{l,m}-1) , &C_m(v_i,v_j,v_k)<\delta, \label{eqn:DMM3} \end{cases} \\
    G_{n,m}(v_n, v_j, v_k) &= \frac{1}{2} q_{n,m} \min[(1-q_{j,m}v_j),(1-q_{k,m}v_k)], \label{eqn:DMM4} \\
    R_{n,m}(v_n,v_j,v_k) &= \begin{cases} \frac{1}{2}q_{n,m}(1-q_{n,m}v_n), &C_m(v_n,v_j,v_k) = \frac{1}{2}(1-q_{n,m}v_n), \\ 0, &\text{else}. \end{cases} \label{eqn:DMM5} \\
    C_m(v_i, v_j, v_k) &= \frac{1}{2}\min[(1-q_{i,m}v_i),(1-q_{j,m}v_j),(1-q_{k,m}v_k)], \label{eqn:DMM6}
\end{align}
\end{widetext}

\noindent where $n = 1, 2, ..., N$, $m = 1, 2, ..., M$, and $q_{i, m} = 1$ if $l_{i, m} = v_i$ (in clause $m$) while $q_{i ,m} = -1$ if $l_{i, m} = \lnot v_i$. Variables are initialized to $v_i(t=0) = \pm 1$ (randomly), $x_m^s(t=0) = 1/2$, and $x_m^l(t=0) = 1$ and bounded so that $v_i \in [-1, 1]$, $x_m^s \in [0, 1]$, and $x_m^l \in [1, 10^4 M]$.

Logical variables are guided to the solution by two competing terms: the ``gradient'' terms $G_{n,m}$ and the ``rigidity'' terms $R_{n,m}$. They are subscripted by $n$, referring to the variable $v_n$ they influence, and $m$, corresponding to a particular clause. If variable $v_n$ does not appear in clause $m$, $G_{n,m} = R_{n,m} = 0$.

Otherwise, the gradient term ``pushes'' $v_n$ so that its corresponding literal evaluates to TRUE ($G_{n,m} > 0$ if $v_n$ is itself the literal, $G_{n,m} < 0$ if $v_n$ is negated), with strength depending on the value of the other two variables in the clause, $v_j$ and $v_k$. If either literal $q_{j,m} v_j$ or $q_{k,m} v_k$ is already close to $1$ (with $1$ corresponding to TRUE, logically), then the magnitude of $G_{n,m}$ will be small. On the other hand, if both literals $q_{j,m} v_j$ and $q_{k,m} v_k$ are near $-1$ (logically FALSE), $|G_{n,m}|$ is larger, as the $q_{n,m} v_n$ needs to be pushed to logical satisfaction (to $1$) to satisfy the clause.

The rigidity term acts differently: it holds a particular $v_n$ at its value if it contributes the most to the satisfaction of clause $m$ (i.e., if it is closer to $q_{n,m}$ than $v_j$ is to $q_{j,m}$ or $v_k$ is to $q_{k,m}$). Otherwise, $R_{n,m}$ does nothing. This asymmetry helps alleviate frustration in the system that would arise from gradient-like terms acting alone, especially for large $N$.

Note that these equations feature {\it two} distinct types of memory DOFs: {\it short-term memories} $x_m^s \in [0, 1]$ and {\it long-term memories} $x_m^l \in [1, 10^4 M]$, both of which act as clause-dependent weights to the gradient and rigidity terms. They both depend on the clause function $C_m(v_i, v_j, v_k) \in [0, 1]$, which quantifies the logical satisfaction of a given clause $m$. Essentially, when $C_m$ is close to $0$, at least one of the literals $q_{i,m} v_{i,m}$ is close to $1$, so the entire clause is satisfied. When $C_m \approx 1$, no literals are TRUE, and the clause is unsatisfied. Intermediate values are, in some sense, ``less'' TRUE or FALSE than the corresponding Boolean ones. However, one can always truncate $v_n$ or $q_{n,m} v_n$ by its sign (positive being TRUE, negative being FALSE) to prescribe an actual Boolean value to any $C_m$. $x_m^s$ keeps track of whether a certain clause $m$ is satisfied by growing when $C_m > \gamma$ and decaying when $C_m < \gamma$. The short-term memories grow faster than $x_m^l$ (we require that $\beta > \alpha$), yet much slower than $v_n$ (necessary to establish DLRO, and to allow the $v_i$'s to settle before perturbing them further). These short-term memories act as ``switches'' between the gradient and rigidity terms; when a clause is unsatisfied, the gradient term dominates, and vice versa. On the other hand, $x_m^l$ tracks the long-term history of a clause's logical satisfaction. Since their upper bound is quite large and $\alpha$, $C_m$, and $\delta$ are all $O(1)$ or smaller, $x_m^l$ can slowly increase for a very long time. If a clause $m$ has been unsatisfied for an extremely long time, $x_m^l$ applies greater pressure to $v_n$, encouraging that clause to become satisfied (after which the long-term memory decays exponentially fast). By keeping both of these variables bounded, the phase space remains compact.

\subsection{Comparison with previous work}

The primary difference between these dynamics and the ones proposed in a prior study~\cite{bearden2020efficient} comes from the memory dynamics. In that work:

\begin{align}
    \dot{x}_m^s &= \beta(x_m^s + \epsilon) (C_m(v_i, v_j, v_k) - \gamma) ,\label{old1}\\
    \dot{x}_m^l &= \alpha (C_m(v_i, v_j, v_k) - \delta) .\label{old2}
\end{align}

Above, $0 < \epsilon \ll 1$ is a small parameter meant to remove spurious solutions when $x_m^s = 0$ ($\epsilon = 10^{-3}$ in~\cite{bearden2020efficient}). Also, the parameters were fixed so that $\{ \alpha, \beta, \gamma, \delta, \zeta \} = \{ 5, 20, 1/4, 1/20, 10^{-3} \}$. No sophisticated tuning procedure was performed, but the chosen parameters still proved quite effective.

With Eq.~(\ref{old1}), the short-term memories $x_m^s$ grow and decay {\it exponentially} rather than linearly (assuming an approximately constant $C_m$ until $v_i$, $v_j$, and/or $v_k$ begin to flip). Instead, by restricting $x_m^s$ so that it does not grow exponentially as in Eq.~(\ref{eqn:DMM2}), $\beta$ is a better indicator of the memory DOFs timescale, allowing us to more effectively maintain timescale separation between the logical and memory DOFs. On the other hand, the long-term memories decay linearly in the former Eq.~(\ref{old2}), whereas in our present model, Eq.~(\ref{eqn:DMM3}), their decay is exponential. This exponential decay assures that clauses $m$ that reached very high $x_m^l$ do not dominate the dynamics for too long after they have become satisfied.

Following some initial testing, we found our updated functional forms to operate better at many points in parameter space relative to those in~\cite{bearden2020efficient} up to moderate $(N \sim 500)$ system sizes. However, which functional form for $\dot{x}_m^s$ or $\dot{x}_m^l$ is most useful remains highly dependent on the parameters chosen and the system size. We expect other functional forms to perform equally well or perhaps even better than Eqns.~(\ref{eqn:DMM1})-~(\ref{eqn:DMM6}) for other types of problems.

\section{Parameter optimization}

There are some physical constraints that the DMM parameters must satisfy which we can use to bound the parameter space (in particular, $\delta$ and $\gamma$). These thresholding parameters satisfy $0 < \delta < \gamma < 1/2$. Both parameters must be less than $1/2$ so that no memories decay while their clause is unsatisfied ($\delta > C_m > 1/2$). On the other hand, it can be advantageous for them to be less than $1/2$ so some memories grow while $C_m$ is already satisfied (this establishes additional ``paths'' in the system through which the logical DOFs can correlate via the memories).

Inspired by the work in Ref.~\cite{10386235} we have used a parallel tempering (PT) approach~\cite{earl2005parallel} for tuning the initial values of $\{\gamma, \delta, \alpha/\beta\}$, and finding the ``optimal'' $\beta$ and $\zeta$ to center our search in the 2D parameter space. We have simulated $R = \lfloor N/20 + 10 \rfloor$ replicas over $1000 / R$ Monte Carlo steps (including proposed PT swaps). Additionally, we have chosen the minimum and maximum replica temperatures $T_{min} = 10^{-4}$ and $T_{max} = 1$ (in units of $k_B = 1$) and space replicas geometrically. We found this spacing, along with this linearly increasing number of replicas, yields a replica swap acceptance probability of approximately 20\% across most system sizes $N$. This is in line with optimal replica temperature distributions in existing literature~\cite{earl2005parallel, 10.1063/1.1917749, 10.1063/1.1831273}.

To fix $T_{min}$ and $T_{max}$, a series of Metropolis updates were performed on 100 isolated replicas (no swapping), each with temperatures from $10^{-7}$ to $10^2$, and the minimum energy $E(T)$ (given by our metric, defined below) found in a fixed number of updates was extracted. After repeating this process many times, we found the profile of the average energy $\langle E(T) \rangle$ flattens around $T_{min} = 10^{-4}$ and $T_{max} = 1$, suggesting there are no features in the energy landscape finer (coarser) that  $T_{min}$ ($T_{max}$) can probe~\cite{earl2005parallel}.

During the PT procedure, we must invoke a particular metric to evaluate the quality of a given set of parameters. We choose to solve instances of size $N$ in sets of three ($N_1$, $N_2$, and $N_3$) and extract the median time-to-solution (TTS) of each, denoted as $m_N$. Then, we can construct the following formula for the metric:

\begin{align}
    \text{metric} &= \text{TTS}_{int} + c_1 \text{TTS}_{slope} + c_2 \text{TTS}_{concav},\\
    \text{TTS}_{int} &= \log_{10}m_{N_1},\\
    \text{TTS}_{slope} &= \frac{\log_{10}m_{N_3} - \log_{10}m_{N_1}}{\log_{10}N_3 - \log_{10}N_1},\\
    \text{TTS}_{concav} &= \frac{\log_{10}m_{N_3} - \log_{10}m_{N_2}}{\log_{10}N_3 - \log_{10}N_2} \\
    & \qquad - \frac{\log_{10}m_{N_2} - \log_{10}m_{N_1}}{\log_{10}N_2 - \log_{10}N_1}.
\end{align}

Essentially, this metric is minimized when the polynomial line of best fit to $\{(N_1, m_{N_1}), (N_2, m_{N_2}), (N_3,  m_{N_3})\}$ (in $\log_{10}$) has a small y-intercept, slope, and concavity (ideally, the concavity will be negative). Heuristically, we find $c_1 = 2$ and $c_2 = 1/2$ to work well. The optimal parameters found at one triple of $N_1$, $N_2$, and $N_3$ are used to initialize the PT procedure at the next triple. As mentioned in the main text, applying this approach to 3-SAT up to $N = 10^2$ provides the following optimal parameters at that size: $\{\alpha/\beta, \beta, \gamma, \delta, \zeta\} \approx \{0.45, 79, 0.36, 0.080, 0.063\}$.

\section{Avalanche definition}

During DMM dynamics, we define avalanches as follows. When any $v_i$ flips sign (i.e., its corresponding boolean value changes), we say an event has occurred at time $t$ and location $i$. This initializes an avalanche of size 1. We then search for similar flipping events at nearest-neighbors of $v_i$, defined as the other logical variables that appear in the same clauses as $v_i$, during the following time $\Delta_{tw}$, called the ``time window''. Within any time $t' \in [t, t + \Delta_{tw})$, if one of the nearest-neighbors of $v_i$ flips, the size of the avalanche is increased by 1, and further events are searched for adjacent to the newest event between times $t'$ and $t' + \Delta_{tw}$. An avalanche stops when no events are found in any of the nearest-neighbors of recent (within a time $\Delta_{tw}$) events in that avalanche. Each logical DOF can only participate once in any avalanche, fixing the maximum possible avalanche size at $N$.

Additionally, individual avalanches are merged when a single $v_i$ flips that is adjacent to at least one event in more than one unique avalanche. In this case, the constituent avalanche sizes are summed into a single new avalanche, and the parent avalanches are discarded. This accounts for the strongly coupled nature of the logical DOFs due to memory; since $v_i$'s that are not nearest-neighbors can significantly impact each others' dynamics, avalanches that begin at multiple locations and later merge must be considered.

\bibliography{supp}